\documentclass[DIV=calc,%
							paper=a4,%
							fontsize=11pt,%
							twocolumn]{scrartcl}	 					

\pdfoutput=1
\pdfmapfile{+txfonts.map}

\usepackage[english]{babel}										
\usepackage[protrusion=true,expansion=true]{microtype}				
\usepackage{amsmath,amsfonts,amsthm}					
\usepackage[final]{graphicx}									
\usepackage{xcolor}									
\usepackage[normal, small, hypcap, up, labelfont=bf, textfont=it]{caption}	
\usepackage{epstopdf}												
\usepackage{subfig}													
\usepackage{booktabs}												
\usepackage{fix-cm}													
\usepackage{amssymb,amsfonts}       
\usepackage{dsfont}
\usepackage{bbm}
\usepackage{pstricks}
\usepackage{cite}
\usepackage[utf8]{inputenc}
\usepackage[perpage,symbol*]{footmisc}
\usepackage{pstricks}
\usepackage[varg]{txfonts}
\usepackage{balance}
\usepackage{fancyhdr}	
\usepackage{hyperref}
\usepackage{verbatim}
\usepackage{fontenc}

\DeclareCaptionFont{mycolor}{\color[HTML]{0000FF}}
\usepackage{sectsty}													
\allsectionsfont{
	\color[HTML]{31ADF3}\usefont{OT1}{phv}{b}{n}
	}

\sectionfont{
	\color[HTML]{31ADF3}\usefont{OT1}{phv}{b}{n}
	}

\usepackage{fancyhdr}												
\usepackage{lettrine}
\newcommand{\initial}[1]{%
     \lettrine[lines=3,lhang=0.3,nindent=0em]{
     				\color[HTML]{31ADF3}
     				{\textsf{#1}}}{}}

\usepackage{titling}															

\newcommand{\HorRule}{\color[HTML]{31ADF3}
									  	\rule{\linewidth}{1pt}%
										}

\pretitle{\vspace{-30pt} \begin{flushleft} \HorRule 
				\fontsize{34}{34} \usefont{OT1}{phv}{b}{n} \color[HTML]{31ADF3} \selectfont 
				}
\title{How Does Nature Accomplish Spooky Action at a Distance?}					
\posttitle{\par\end{flushleft}\vskip 0.5em}

\preauthor{\begin{flushleft}\large \lineskip 0.5em \usefont{OT1}{phv}{b}{sl} \color[HTML]{31ADF3}}
\author{Mani L. Bhaumik\\[8pt]}											
\postauthor{\footnotesize \usefont{OT1}{phv}{m}{sl} \color[HTML]{000000} 
Department of Physics and Astronomy, University of California, Los
Angeles, USA. E-mail: \href{mailto:bhaumik@physics.ucla.edu}{bhaumik@physics.ucla.edu}\\[10pt]		
\scriptsize\usefont{OT1}{phv}{m}{n} \color[HTML]{31ADF3}{\textbf{Editors: \emph{Zvi Bern} \& \emph{Danko Georgiev}} }\\[5pt]
\color[HTML]{000000}{Article history: Submitted on November 26, 2018;  Accepted on December 21, 2018; Published on December 23, 2018.}
					\par\end{flushleft}\HorRule}

\date{}																				

\begin{document}
\maketitle
\thispagestyle{fancy} 			
\initial{T}\textbf{he enigmatic nonlocal quantum correlation that was famously derided
by Einstein as ``spooky action at a distance'' has now been experimentally demonstrated
to be authentic. The quantum
entanglement and nonlocal correlations emerged as inevitable
consequences of John Bell's epochal paper on Bell's inequality.
However, in spite of some extraordinary applications as well as attempts
to explain the reason for quantum nonlocality, a satisfactory account of
how Nature accomplishes this astounding phenomenon is yet to emerge.
A cogent mechanism for the occurrence of this incredible event is
presented in terms of a plausible quantum mechanical Einstein--Rosen
bridge.\\ Quanta 2018; 7: 111--117.}
\begin{figure}[b!]
\rule{245 pt}{0.5 pt}\\[3pt]
\raisebox{-0.2\height}{\includegraphics[width=5mm]{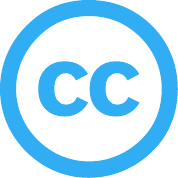}}\raisebox{-0.2\height}{\includegraphics[width=5mm]{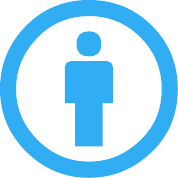}}
\footnotesize{This is an open access article distributed under the terms of the Creative Commons Attribution License \href{http://creativecommons.org/licenses/by/3.0/}{CC-BY-3.0}, which permits unrestricted use, distribution, and reproduction in any medium, provided the original author and source are credited.}
\end{figure}
\section{Introduction}

In 1975 a young French doctoral candidate Alain Aspect went to John
Bell's office at CERN seeking his advice about the suitability of
carrying out experiments on Bell's inequality for a thesis. After listening
to the proposal, the first question from Bell was, 
\begin{quote}
``Have you a permanent position?'' \cite[p.~119]{Aspect2002}
\end{quote}
Obviously, little did Bell know at the time that the
consequences of his rather modest paper \cite{Bell1964} in an obscure, short lived
journal would explode in bringing about one of the most profound
discoveries of Nature.

For more than a decade, John Bell has earned his living working almost
exclusively on theoretical particle physics and accelerator design at
CERN. But his prodigious curiosity drove him to devote much of his
thinking to a rigorously honest exploration of how Nature really works,
a pursuit which Einstein had considered being the heart of science. From
his own experience as a student, he felt that generations of
undergraduates in quantum physics had been steered away from thinking
too deeply about the reality of quantum physics and thus effectively
brainwashed. This conviction drew Bell to the foundational aspects of
quantum physics.

In 1964, after a year's leave from CERN, spent at various universities in
the United States, he published his seminal paper what is now known as
the famous Bell's theorem in the now defunct journal, Physics. With his
exceptional keen insight, Bell devised for the first time an experimental
procedure \cite{Bell1964} to decipher the long standing Einstein--Podolsky--Rosen
(EPR) paradox \cite{Einstein1935a}, which had been sadly ignored for three decades as
merely a philosophical quandary. Bell, a devotee of Einstein and an
admirer of Bohm, had set out initially to prove Einstein right in his
dismissal of ``spooky action,'' and in search of the Bohm--de Broglie
hidden variables. But science does not always take us where we expect
to be taken, and he wound up with a very different conclusion.

For the proposed experiments, he devised a formulation known as Bell's
inequality. Experimental results violating Bell's inequality will prove
the predictions of quantum mechanics known as entanglement and
nonlocality to be true. What Bell found was that there was no theory of
local hidden variables that could account for these predictions. For a
decade, this remarkable discovery was essentially ignored. No one
envisioned that Bell's theorem will bring about an inconceivable
transformation in physics. Further details on quantum entanglement have been presented in
a previous article by Bhaumik \cite[\S5]{Bhaumik2017}.

The very first indication of the violation of Bell's inequality was
presented in 1972 by Stuart J. Freedman and John F. Clauser \cite{Freedman1972}. The
first convincing experiment was carried out by Alan Aspect and his
colleagues \cite{Aspect1982}. Countless papers followed, closing nearly all the possible
loopholes in the theory as well as experiment. But for a minority of
hold-outs, quantum entanglement and nonlocality are now generally
accepted as genuine phenomena of Nature despite their strikingly
counter intuitive characteristics. These are discussed in an abundant
number of articles. A representative samples may be found in the works of Ryszard
Horodecki and colleagues \cite{Horodecki2009} and Dominik Rauch and colleagues \cite{Rauch2018}.

Successful experiments validate theories, but it is the application of
the findings that leads to widespread acceptance of a new truth. For
many years, entanglement and nonlocality were treated largely as mere
philosophical matters by much of the physics community. However,
since the first usage \cite{Ekert1991} in secure, quantum cryptography and the
possibility of devising a quantum computer that could be a trillion times faster than the fastest digital super computer, the field has
literally exploded. Countless research articles published in the most
prestigious scientific journals are effectively erasing any trace of
skepticism.

Applications of quantum entanglement and nonlocality span a wide and
varied range: from quantum cryptography \cite{Bennett2014,ShenoyHejamadi2017} to secure
communication systems, and even a global quantum internet. All are
now deemed possible. Quantum teleportation has also been
demonstrated. But perhaps the most promising application could be in
the development of quantum computers, which is the object of feverish
research worldwide.

Entanglement is now finding applications in very diverse fields. In
defense application, development of a radar system that uses quantum
entanglement to beat the stealth technology of modern military aircraft is
being considered \cite{LasHeras2017}. Presumably the most promising resolution of the
quantum measurement problem at the moment appears to be the theory
of decoherence, which is intimately connected with entanglement. Some
eminent scientists even consider spacetime itself to be stitched together
with quantum entanglement.

Heaps of praise is now being bestowed on the erstwhile unheeded Bell's
theorem to be one of the most ingenious discoveries of science.
\begin{quote}
``In 1964, John Bell fundamentally changed the way that we think about
quantum theory,'' \cite{Leifer2014}
\end{quote}
pronounces Mathew S. Leifer. Alain Aspect adds
grandly,
\begin{quote}
``I think it is not an exaggeration to say that the realization of
the importance of entanglement and the clarification of the quantum
description of single objects have been at the root of a second quantum
revolution, and that John Bell was its prophet.'' \cite{Aspect2004}
\end{quote}

\section{Quantum Entanglement}

Although quantum entanglement has been demonstrated experimentally
with photons \cite{Kwiat1995,Zhao2004,Lu2007,Yao2012}, neutrinos \cite{Formaggio2016}, electrons \cite{Hensen2015}, molecules as
large as Bucky balls \cite{Arndt1999,Nairz2003} and even small diamonds \cite{Lee2011}, by far
the most expedient way to produce and study entanglement is using
polarized photons. In many experimental studies involving some
application of entanglement, polarized photons are used as quantum
particles since they are much easier to handle and preserve their
coherence over a long distance, especially in passage through air, which
is not dichroic.

In the exceptionally popular procedure \cite{Kwiat1995}, the desired polarization-entangled
states are produced directly out of a single nonlinear BBO
crystal. By means of such a system, one can very easily produce any of
the four maximally entangled EPR-Bell states of the two photons,
\begin{eqnarray}
|\Psi^{\pm}\rangle	& = &	\frac{1}{\sqrt{2}}\left(|H\rangle_{1}|V\rangle_{2}\pm|V\rangle_{1}|H\rangle_{2}\right),\nonumber \\ 
|\Phi^{\pm}\rangle	& = &	\frac{1}{\sqrt{2}}\left(|H\rangle_{1}|H\rangle_{2}\pm|V\rangle_{1}|V\rangle_{2}\right),
\label{eq:1}
\end{eqnarray}
where $|H\rangle_{1}$, $|V\rangle_{1}$ denote the horizontal and vertical polarization state of
photon 1 and $|H\rangle_{2}$, $|V\rangle_{2}$ represent the same aspects of photon 2.

The special property of nonlocality of quantum entanglement is
observed when we separate the two photons by an arbitrarily large
distance. Now, if we measure the polarization of one of the photons, get
a result, and then measure the polarization of the other photon along the
same axis, we find that the result for the second photon is correlated.
The wave function of the second photon as well as the probability
distribution for the outcome of a measurement of the polarization along
any of its axis changes upon measurement of the first photon. This
probability distribution is in general different from what it would be
without measurement of the first photon.

Rather surprisingly a photon can even be entangled with another one
created even at a subsequent time. E.~Megidish and colleagues \cite{Megidish2013} created a
photon pair (1-2). After measuring photon 1, the second photon can be
stored for example in an optical delay line while a second pair (3-4) is
created. Photons 2 and 3 are then projected onto the Bell basis, which
swaps entanglement onto photons 1 and 4. Since photon 1 and 4 display
quantum correlation even though they never coexisted, entanglement not
only holds for space like separation but for time like separation as well.
Incidentally, entanglement swapping plays a crucial role in quantum
repeaters essential for overcoming loss of photons in long distance
quantum communications.

Although so far we have discussed a bipartite state consisting of photons
only, Eq.~\eqref{eq:1} holds in general for any two entangled qubits A and B
each in an incoherent superposition of $|0\rangle$ and $|1\rangle$ \cite[\S4.17]{Georgiev2017}.
The four maximally entangled pure Bell states for two qubits are usually
given by,
\begin{eqnarray}
|\Psi^{\pm}\rangle	& = &	\frac{1}{\sqrt{2}}\left(|0\rangle_{A}|1\rangle_{B}\pm|1\rangle_{A}|0\rangle_{B}\right),\nonumber \\ 
|\Phi^{\pm}\rangle	& = &	\frac{1}{\sqrt{2}}\left(|0\rangle_{A}|0\rangle_{B}\pm|1\rangle_{A}|1\rangle_{B}\right).
\end{eqnarray}
These states form an orthonormal basis of the Hilbert space of the two
qubits.

In order to illustrate nonlocal correlation, let us choose the $|\Phi^{+}\rangle$ state
where there is equal probability of measuring either product state
$|0\rangle_A|0\rangle_B$ or $|1\rangle_A|1\rangle_B$ as $|\frac{1}{\sqrt{2}}|^2 = \frac{1}{2}$.

Now let us separate the two entangled qubits by an arbitrarily large
distance and give one qubit each to Alice and Bob. If Alice makes a
measurement of her qubit, she will get either $|0\rangle_A$ or $|1\rangle_A$ with equal
probability and cannot tell if her qubit had value 0 or 1 because of the
qubit's entanglement. If Bob now measures his qubit, he must get
exactly the same result of his measurement as Alice. For example, if Alice
measures $|0\rangle_A$, Bob must measure $|0\rangle_B$ since $|0\rangle_A|0\rangle_B$ is the only
state in $|\Phi^{+}\rangle$ where Alice's qubit is $|0\rangle_A$. So, for these two entangled qubits,
whatever Alice measures, so would Bob, with perfect correlation,
however far apart they may be and even though neither can predict in advance
whether their qubit would have value of 0 or 1.

This conclusion, by itself, is compatible with an interpretation in
which the qubits have definite values before they are measured, with
probabilities arising simply due to our ignorance of these values.
However, by considering further types of measurements that project
onto linear combinations of $|0\rangle$ and $|1\rangle$, Bell showed that such an
interpretation cannot account for the correlations that quantum
mechanics predicts, without assuming some instantaneous action at a
distance.

\section{No Violation of Relativity}

Although the enchanting ``spooky action at a distance'' has been
discovered to be real, which would have been a cause of
great consternation to Einstein, he would have been very happy to know
that it does not violate his cherished special theory of relativity. Because
no useful information can be transmitted instantaneously using quantum
nonlocality.

This is due to the fact that the actions of an experimentalist on a
subsystem of an entangled state can be described as applying a unitary
operator to that subsystem. Although this produces a change on the wave
function of the complete system, such a unitary operator cannot change
the density matrix describing the rest of the system. In brief, if
distant particles 1 and 2 are in an entangled state, nothing an
experimentalist with access only to particle 1 can do that would change
the density matrix of particle 2.

The density matrix $\hat{\rho}$ of an ensemble of states $|n\rangle$ with probabilities $P_n$ is given by
\begin{equation}
\hat{\rho}=\sum_{n}P_{n}|n\rangle\langle n|,
\end{equation}
where $|n\rangle\langle n|$ are projection operators and the sum of the probabilities is
$\sum_{n}P_{n}= 1$. Thus there can be various ensembles of states with each one
having its own probability distribution that will give the same density
matrix.

The mean value of an observable $\langle\hat{A}\rangle$ is:
\begin{equation}
\langle\hat{A}\rangle=\textrm{Tr}\left(\hat{\rho}\hat{A}\right).
\end{equation}
Furthermore, the time evolution of the density matrix $\hat{\rho}(t)$ only depends upon the commutator $[\hat{H},\hat{\rho}(t)]$ following
the von Neumann equation,
\begin{equation}
i \hbar \frac{\partial}{\partial t} \hat{\rho} (t) = [\hat{H},\hat{\rho}(t)],
\end{equation}
where $\hat{H}$ is a Hermitian operator called the Hamiltonian.

Thus as long as $\hat{\rho}$ remains the same, a change in the wave function of
particle 2 does not affect any observable since all observable results can
be predicted from the density matrix without needing to know the
ensemble used to construct it. Consequently, no useful signal can be sent
using entanglement and nonlocality between two observers separated by
an arbitrary distance thereby no violation of the sanctified tenets of
special theory of relativity ensues.

Still there are some very valuable applications that can be realized for
example in secure quantum cryptography and communication system
where the system destroys itself when an intruder eavesdrops.

\section{How does Nature accomplish nonlocality}

Apart from any possible practical application, it would be thrilling to
uncover how such an astonishing event can take place in Nature. On this
point, even Einstein would enthusiastically concur, since underlying his
debates with Bohr was his contention that science should seek to explain
how Nature works, not simply to tell us what we can know about how it
works. A sketch of a possible exploratory process using the
characteristic fluctuations of the quantum fields to form an Einstein--Rosen (ER) bridge \cite{Einstein1935b} was presented by the author in a previous
publication \cite{Bhaumik2014}. Some further elaboration of the possibility will now be
presented. In order to accomplish that, it would be necessary to clearly
understand the inherent, inseparable connections of the entangled
particles with the underlying quantum fields and their vacuum
fluctuations.

By way of many experiments over the years, the Quantum Field Theory
of Standard Model has successfully explained almost all experimental
observations in particle physics and correctly predicted a wide
assortment of phenomena with impeccable precision and according to
the experts, QFT is here to stay as an effective field theory.

Steven Weinberg asserts,
\begin{quote}
``the Standard Model provides a remarkably unified view of all types of matter and forces (except for
gravitation) that we encounter in our laboratories, in a set of equations that can fit on a single sheet of paper. We can be certain
that the Standard Model will appear as at least an approximate feature of
any better future theory.'' \cite[p.~264]{Weinberg2015}
\end{quote}
And Frank Wilczek affirms,
\begin{quote}
``the standard
model is very successful in describing reality---the reality we find
ourselves inhabiting.'' \cite[p.~96]{Wilczek2008}
\end{quote}

Wilczek additionally enumerates the most crucial aspects of the quantum
fields as the primary constituents of everything physical in this universe.
\begin{quote}
``The primary ingredient of physical reality, from which all else is formed, fills all space and time.
Every fragment, each space-time element, has the same basic properties as every other fragment.
The primary ingredient of reality is alive with quantum activity.
Quantum activity has special characteristics. It is spontaneous and unpredictable.''
\cite[p.~74]{Wilczek2008}.
\end{quote}
He continues further to pronounce, 
\begin{quote}
``The deeper properties of quantum field theory, which will form the subject of
the remainder of this paper, arise from the need to introduce \emph{infinitely
many degrees of freedom}, and the possibility that all these degrees of
freedom are excited as quantum mechanical fluctuations.'' \cite[pp.~338--339]{Wilczek2006}
\end{quote}
Furthermore,
\begin{quote}
``Loosely speaking, energy can be borrowed to make evanescent virtual
particles. Each pair passes away soon after it comes into being, but new
pairs are constantly boiling up, to establish an equilibrium
distribution.'' \cite[p.~404]{Wilczek2006}
\end{quote}

According to QFT, a particle like an electron is a propagating ripple
(quantized wave) of the underlying electron quantum field, which acts as
a particle because of its well-defined energy, momentum, mass, charge,
and spin, which are conserved properties of the electron \cite{Bhaumik2016}.

Since electrons carry electric charge, their very presence disturbs the
electromagnetic field around them.
The disturbance in the photon or the
electromagnetic field in turn can cause disturbances in other electrically
charged quantum fields, like the muon and the various quark fields.
Generally speaking, in this manner, every quantum particle spends some
time as a mixture of other particles in all possible ways. However, the
combination of the disturbances in the electron field together with those
in all the other fields always maintains well-defined conserved
quantities.

All these disturbances distort the shape of the ripple. However,
irrespective of that shape, it can be expressed as a wave packet by
Fourier analysis. Thus a wave packet is a holistic ensemble of
disturbances of the primary reality of quantum fields, only the totality of
which represent a particle like an electron or a photon and therefore
always needs to be treated as such. It should now be abundantly clear
how intimately and inseparably the quantum particles like photons are
always connected to quantum fields and their inherent fluctuations.

Despite the roiling ocean of quantum fluctuations, some order can be
found in the midst of all the unpredictability. A familiar example is the
decay of radioactive atoms. The instance of decay for any particular
atom is completely spontaneous and totally unpredictable. But for a
sufficient number of these atoms, the time required for the decay of half
of them is evidently calculable. Likewise, the random quantum
fluctuations of the fields in any spacetime element can be embodied in a
wave function (quantum state).

To a good approximation, in QFT the wave function of quantum
fluctuations can be represented by a linear superposition of harmonic
oscillator wave functions. The wave function $|\Psi\rangle$ of quantum
fluctuations in any element of spacetime can be written as a vector in
Hilbert space,
\begin{equation}
|\Psi\rangle=\sum_{n}c_{n}|\Psi_{n}\rangle
\end{equation}
where $|\Psi\rangle$ is normalized so that $\langle |\Psi|\Psi\rangle= 1$.

As ascertained from relativistic Lorentz invariance, any quantum field is
immutable at each space-time element throughout the universe. This
holds in spite of the infinite degrees of freedom for creation of the
vacuum fluctuations at all spacetime elements where any particular
fluctuation is totally spontaneous and completely unpredictable as to
exactly when that fluctuation will take place. Therefore $|\Psi\rangle$ represents
an irreducible randomness that manifests itself throughout the universe
without propagating from one point of space to another.

Since the expectation value of an underlying quantum field in its ground
state is the same throughout the universe, reflecting this reality the state
vector $|\Psi\rangle$ representing its vacuum fluctuations in any spacetime
element should be the same all over the universe, resulting in a
stupendous ensemble of identical quantum states. Because of the
plethora of interactions between the quantum fields predicted by QFT,
there will be entanglement \cite{Narnhofer2012} between all the $|\Psi\rangle$ throughout the
universe. To give just one example, the gravitational field will interact
with all the fields. Of course, the number of interactions that contribute
to various degrees of entanglement on a universal scale is beyond listing.
Consequently, it should be possible to construct a universal Einstein--Rosen (ER) bridge
comprising the entangled $|\Psi\rangle$ states of all the space
time elements.

As discussed earlier, a wave packet representing a quantum particle is a
holistic ensemble of disturbances of physically real quantum fields, only
the totality of which represents a particle like an electron or a photon.
Thus any photon is always entangled with the vacuum and its
fluctuations and consequently the ER bridge. Such a possibility seems to
have been demonstrated experimentally \cite{Fuwa2015}.

When two photons are created simultaneously by down conversion in a
nonlinear crystal, both of them will be entangled with the ER bridge.
However, those photons, which are also in a maximally entangled Bell
state will be concurrently entangled with the vacuum state ER bridge
from their very inception thus causing the bridge to be maximally
entangled as well. This is because the entanglement of the ER bridge with the
maximally entangled photons is stronger compared to when the photons
are not entangled. The monogamy of entanglement is not violated in this
case since the two entanglements are created simultaneously and there is
no cloning.

Thus a possible mechanism for a ``spooky action at a distance'' can be
envisioned. Hrant Gharibyan and Robert F. Penna \cite{Gharibyan2014} have mentioned
that classical ER requires monogamous EPR, stating also that quantum
ER bridges have yet to be defined independently of the ER=EPR
conjecture \cite{Maldacena2013,Cowen2015,Susskind2016a,Susskind2016b}. We believe the possibility of a quantum ER bridge
presented here is worth further investigation. Extended rigorous analysis
including entropy of entanglement as its measure should be carried out.
The maximally entangled ER bridge can facilitate the two maximally
entangled photons to maintain their nonlocal correlations even when
they are separated by an arbitrary distance.

\section*{Acknowledgement}

I would like to thank Zvi Bern, Per Kraus and Danko Georgiev for helpful discussions and valuable comments.

\balance

\end{document}